\documentclass[preprint, 12pt]{elsarticle}

\usepackage[colorlinks=true, linkcolor=black, citecolor=blue, urlcolor=blue]{hyperref}

\usepackage{amssymb}
\usepackage[T1]{fontenc}
\usepackage{pdflscape}
\usepackage{longtable}
\usepackage{hanging}
\usepackage{lineno}
\usepackage{multirow}
\usepackage{array}
\usepackage{arydshln}
\usepackage{threeparttable}
\usepackage{orcidlink}

\journal{Journal of Quantitative Spectroscopy \& Radiative Transfer}

\begin{document}
\begin{frontmatter}

\title{Spectrum of five-times ionized krypton: Kr VI}
\author[a]{Aftab Alam}\affiliation[a]{organization={Department of Physics}, addressline={Aligarh Muslim University}, city={Aligarh}, postcode={202002}, state={Uttar Pradesh}, country={India}}
\author[a,b]{Abid Husain}%\orcidlink{0000-0002-0896-9719}}
\affiliation[b]{organization={Department of Applied Physics}, addressline={Mahatma Jyotiba Phule Rohilkhand University}, city={Bareilly}, postcode={243006}, state={Uttar Pradesh}, country={India}}
\author[a]{K. Haris \orcidlink{0000-0002-1341-6297}\corref{*}}
\author[a]{A. Tauheed} %\orcidlink{0000-0003-3856-5042}}
\author[a]{S. Jabeen}

\cortext[*]{Corresponding author: kharisphy@gmail.com, kharis.ph@amu.ac.in}

\begin{abstract} 
This work describes the spectral analysis of five-times ionized krypton ion (Kr VI) using a high-resolution spectrogram recorded on a 3 m normal incidence vacuum spectrograph in the wavelength 230--2075~\AA~region. For spectral excitation, a gas-puff triggered spark source was used. This work thoroughly examined all previously reported spectroscopic analyses for Kr VI. Many missing and new levels were added to the list of known energy levels with the help of several supportive transitions. The present experimental findings were theoretically supported within the pseudo-relativistic Hartree-Fock (HFR) formalism implemented in the Cowan suite of codes. A total of 52 (including eight new) energy levels were established with the help of 169 unique observed spectral lines (31 are new) assigned to 175 (6 doubly assigned) transitions. All observed and Ritz wavelengths were reported with their uncertainties, modeled intensities, and other evaluated radiative transition parameters such as transition probabilities and cancellation factors.
\end{abstract} 
 
%%Graphical abstract
%\begin{graphicalabstract}
%\includegraphics{grabs}
%\end{graphicalabstract}

%%Research highlights
% \begin{highlights}
% \item Spectroscopic analysis of five-times ionized krypton (Kr VI).
% \item Accurate energy levels with their uncertainties, observed and Ritz wavelengths with their uncertainties, and transition probabilities are presented for the Kr VI spectrum.
% \item Theoretical interpretations for the present findings were made using the Hartree-Fock method with relativistic corrections (HFR) implemented in Cowan's codes.
% \end{highlights}

\begin{keyword}
Atomic spectra, Ionized krypton, Energy levels, Spectral lines, Wavelengths, Transition probabilities.
\end{keyword}
\end{frontmatter}
%% main text

\section{Introduction} \label{sec:intro}
 
The spectral transitions of krypton ions have been marked as an important species for nebular spectroscopy, understanding nucleosynthesis of elements beyond the iron peak (Z > 30) by neutron capture process, particularly to study the s-process enrichment in several astrophysical nebulae \citep{Pequignot_1994_appl, Dinerstein_2001_Se4_Kr3, Sterling_2017_Kr6_PNe}. Recently, fourteen spectral lines of five-times-ionized krypton (Kr VI) were detected in the UV spectrum of a white dwarf RE 0503--289 (hot DO-type) \citep{Werner_2012, Rauch_2016_Kr-VI}. Krypton was injected into fusion reactors, such as ITER, as a potential W-equivalent radiator \citep{Turco_2024_Kr_Xe_in_ITER}; therefore, its accurate and precise spectral data play a significant role in plasma diagnostics. \\

The Kr VI is a gallium (Ga I) isoelectronic sequence member. It has a ground electronic configuration 4s$^{2}$4p with $^2$P$^{\circ}_{1/2, 3/2}$ levels. The excited levels are produced either by single-electron configuration types 4s$^{2}n\ell$~($n \ge 4$, $\ell = s, p, d,~....$) or by three-electron configurations such as 4s4p$^{2}$, 4p$^3$, 4p$^2$4d, and 4s4p($^{1, 3}P^{\circ}$)$n\ell$~($n \ge 4$, $\ell = s, p, d,~....$); thus, its spectroscopic term structure comprises of doublets and quarters manifold. For the first time, the spectrum of Kr VI was reported by \citet{Fawcett1961-VUV-inert-gas} using a zeta-pinch light source, followed by its capillary discharge (CD) spectrum by \citet{schonheit_1965_spektren}. Many previous beam-foil spectroscopic studies reported a few lines of Kr VI \citep{Druetta_1976_Kr-VI, Irwin_1976_Kr_BFS, Livingston_1976_Kr4-7}. In 1988, \citet{Trigueiros_1988_Kr6} reported additional lines of Kr VI from a theta-pinch source. In the following years, \citet{Tauheed_1990_Kr-VI} identified more new lines, including the inter-combination lines 4s4p~$^2$P$^{\circ}$--4s4p$^2$~$^4$P of Kr VI using the beam-foil technique. The extended analysis of Kr VI was provided by Pagan and his colleagues \citep{Pagan_1995_Kr6_exp, Pagan_1996_Kr6_os} by taking its multiple spectra using theta-pinch and CD light sources. A few Rydberg lines of Kr VI were observed in different collisional experiments \citep{Martin_1993_Krq+-Kr_colli, Jacquet_1994_Kr8-Li_coll, Chen_2001_Kr_ebit, Chen_2002_Kr_ebit}. In 2007, \citet{Saloman_2007_Kr_compil} explored all these spectral data for his compilation work on krypton spectra; accordingly, he generated the optimized energy levels for the spectrum from the best available observed wavelengths. His work superseded all previous compilations in the same regard (see details in ref. \citep{Saloman_2007_Kr_compil}). After Saloman's compilation, \citet{Farias_2011_Kr-VI} established the energy levels of 4s4p5p with a few known levels of 4p$^2$4d and 4s4p4f configurations. In many other experiments, the Kr VI lines were observed \citep{Lu_2006_Kr6_photo-ex-ion, Lundeen_2007_Kr6+_polariz, Knie_2012_EIE_noble_gas}, and some confirmed the previous identifications \citep{Pagan_1995_Kr6_exp, Pagan_1996_Kr6_os}. In Kr VI, the experimentally measured lifetime data are available for a limited number of low-lying excited levels \citep{Irwin_1976_Kr_BFS, Livingston_1976_Kr4-7, Tauheed_1990_Kr-VI}, and no measurements exist for other radiative parameters, such as transition probabilities (TPs or gA-values) and oscillator strengths, except those in ref.~\citep{Lu_2006_Kr6_photo-ex-ion}. On the other hand, several theoretical calculations were found in the literature~\citep{Rauch_2016_Kr-VI, Pagan_1996_Kr6_os, Lu_2006_Kr6_photo-ex-ion, Aashamar_1983_os_th_ga-seq, Biemont_1987_tp_fb, Biemont_1990_os_ga-like, Marcinek_1993_os_ga-like, Biemont_1994_Comm_os_ga-like, Ali_1997_fine-str_Ga-like, Safronova_2006_4p_Ga-like, Charro_2008_tp_Ga-like, Hu_2011_tp_Ga-like}. Only two of them -- Pagan et al. \citep{Pagan_1996_Kr6_os} and \citet{Rauch_2016_Kr-VI} -- are noteworthy as they provided extensive data sets, which would be helpful for us to perform the TP comparison and evaluation.\\

In the present study, we intend to (re-)investigate the spectrum of Kr VI with the help of the krypton spectra recorded on a 3~m normal incidence vacuum spectrograph (NIVS) and also by giving comprehensive theoretical support for our findings with the pseudo-relativistic Hartree–Fock (HFR) approach. We aim to provide the evaluated spectral data, such as energy levels, observed and Ritz wavelengths, transition probabilities, modeled line intensities, and optimized energy levels with their uncertainties for the Kr VI spectrum. \\

\section{Experimental details} \label{sec:exp_detail}

The spectral plates of krypton used in this study were recorded in the~{230--2075~\AA} wavelength range using a 3 m NIVS at the Antigonish laboratory in Canada. The spectrograph was equipped with a holographic grating of 2400 lines per millimeter, providing a reciprocal dispersion of about 1.38 \AA~mm$^{-1}$ in the first diffraction order. A vacuum spark discharge containing carbon and/or aluminum electrodes was used as a light source. To achieve the ionic spectra of krypton, a puff of gas was injected (through the cathode) into the gap between the electrodes to initiate the electrical discharge in a vacuum. The gas puff triggers the spark between the electrodes and discharges a low-inductance capacitor of 14 $\mu${F} (charged up to 6 kV) connected in series with the electrodes. The spectrum produced with carbon electrodes was steady, and the excitation conditions were not much affected during the experiment. In contrast, those with aluminum electrodes were altered by variation of excitation conditions due to the deformation of the electrodes. The exposures were taken on Kodak short-wave radiation (SWR) and 101-05 photographic plates. Multiple exposures, up to five, in different spectral tracks were captured under varying experimental conditions: either by reducing discharge voltage or inserting inductor coils in series with the spark circuit. Further details of the experimental setup are available in ref. \citep{Tauheed_1993_Xe5}. The variation in line intensity of the spectral lines at different spectral tracks corresponding to different discharge settings allowed us to achieve satisfactory ionization separation. The track with the strongest Kr VI lines was opted for measuring the wavelengths, and the remaining tracks on each plate were used to accomplish the ionization separation. The line positions of spectrograms were measured on a Zeiss Abbe comparator at Aligarh. The spectrograms were calibrated using reference lines of C I-IV, N I-II, O I-V, Al II-V, and Si II-IV, together with the selected strong lines of Kr II, whose Ritz wavelengths were taken from the NIST ASD~\citep{nist}. The general estimate of wavelength uncertainty is due to the combined effect of the statistical deviation of the line position measured on the comparator and the systematic uncertainty of the reference wavelengths used in the fitting \citep{Haris2014-Sn2}. The final wavelength uncertainty for sharp and unperturbed lines is estimated to be 0.008~{\AA}, and the given figure was doubled for all perturbed lines. \\

The spectrogram contains krypton lines from several charge states (Kr II--IX). We accessed their spectral data, including observed and Ritz wavelengths, from the NIST ASD~\citep{nist}. It should be noted that a major chunk of spectral data on Kr ions was due to the joint efforts of spectroscopy groups in Argentina and Brazil (Pagan 2025, Private communication; see additional information in refs. \citep{Saloman_2007_Kr_compil, Almandos_2017_review_of_rare-gas_spectra}). Thus, our newly calibrated wavelengths were compared with those previously reported wavelengths, including those of Kr VI by Pagan et al. \citep{Pagan_1995_Kr6_exp, Pagan_1996_Kr6_os}. No discernible systematic differences were found between these measurements when compared based on the wavelength scale. However, it turns out that the quoted measurement uncertainty (a general estimate) of Pagan et al. is insufficient to derive the accurate energy levels for Kr VI. Further discussions are given in Section \ref{sec:lev_opt}. \\

\section{Results and~discussion} \label{sec:result_dis}

\begin{figure*}[ht!]
\centering
\includegraphics[scale=0.5]{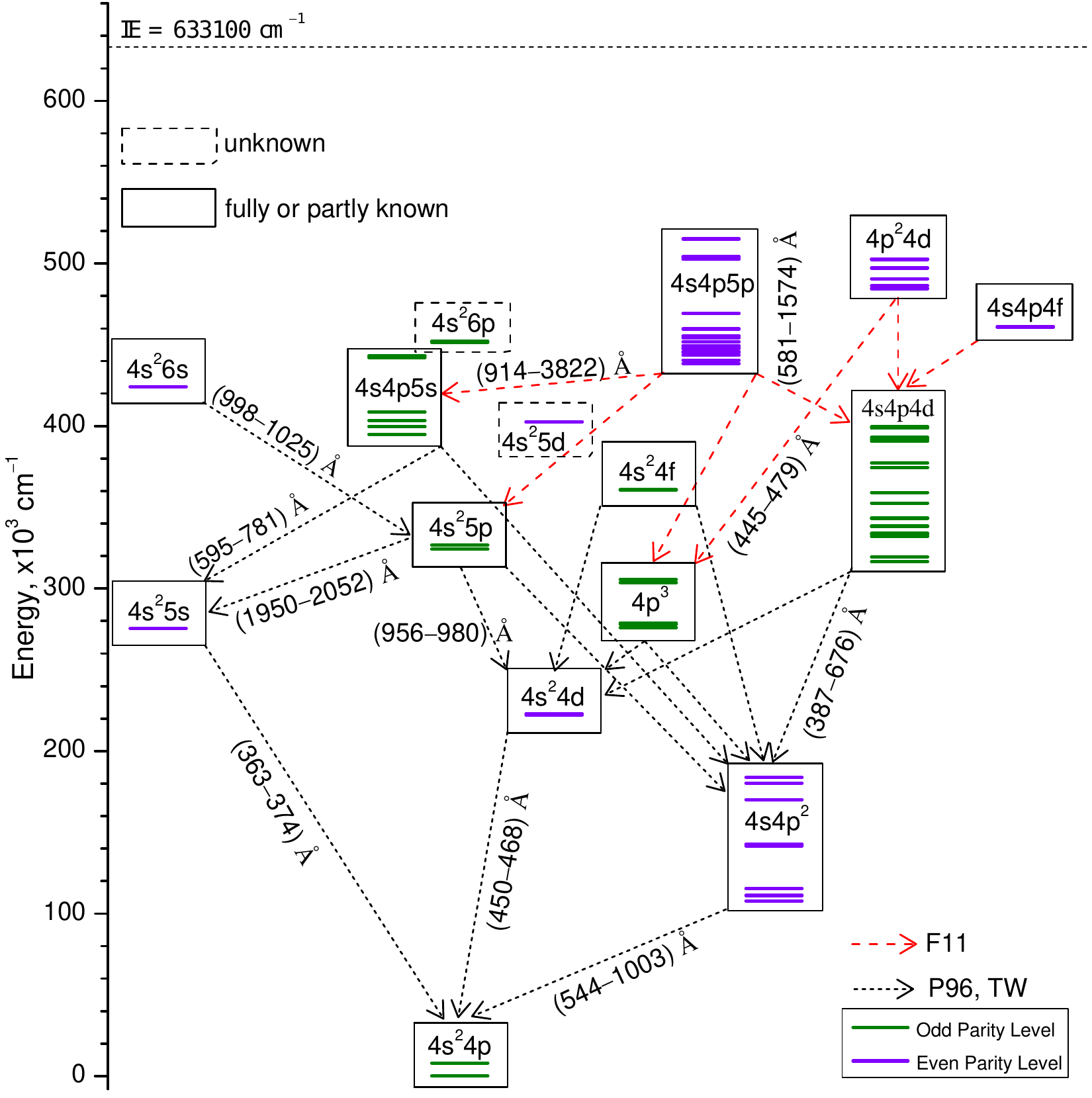}
\caption{Grotrian diagram for known configurations of Kr VI. Its prominent transition arrays are shown with their wavelength ranges, which were observed by P96---Pagan et al. \citep{Pagan_1995_Kr6_exp, Pagan_1996_Kr6_os}, F11---\citet{Farias_2011_Kr-VI}, and TW---this work. The ionization energy (IE) is taken from NIST ASD~\citep{nist}.} \label{fig:energy_lev}
\end{figure*}

The main results of the present study are given in the form of two tables: Table \ref{tab:lin_KrVI} contains the list of classified lines, and Table \ref{tab:lev_KrVI} describes the optimized energy levels with their LS compositions for the Kr VI spectrum. The present experimental findings were supported within an extensive HFR calculation framework with least-squares fit (LSF) adjustments, followed by the evaluation of transition probabilities for the Kr VI spectrum (see Section \ref{sec:theo_HFR}). A Grotrian energy level diagram of Kr VI is given in Figure \ref{fig:energy_lev}. Apart from illustrating the experimentally studied configurations with their known energy levels, this diagram provides insight into interacting configurations that result in the complex term structure for Kr VI. \\ 

%%%%%%%%%%%%%%%%%%%%%%%%%%%%%%%%%%%%
%%%%%%%%%%%% Tables---Lines %%%%%%%%%%%%%
%%%%%%%%%%%%%%%%%%%%%%%%%%%%%%%%%%%%%%%%%%
%\input{Kr6_table1}

\setlength\LTleft{-2cm}
\setlength\LTright{-2cm}
\begin{landscape}
\advance\leftskip-3cm
\advance\rightskip-3cm
\noindent \vspace{-2cm}
\begin{scriptsize}
\begin{longtable}{lccccccccccccccll} 
\caption{Classified lines of Kr VI}
\label{tab:lin_KrVI}
\\
\hline
\hline
I$_{Obs.}${$^{a}$} & $\lambda_{Obs.}${$^{b}$} & Unc.{$^{b}$} & \multicolumn{2}{c}{Classification{$^{c}$}} & $\lambda_{Ritz}${$^{b}$} & Unc.{$^{b}$} & $\delta\lambda_{O-R}${$^{d}$} & gA$_{TW}${$^e$} & |CF|{$^e$} & Acc.{$^f$} & gA$_{R16}${$^g$} & Line Ref.{$^h$} & Comm.{$^i$} \\
(arb.~u.)& (\AA) & (\AA) & Lower level & Upper level & (\AA) & (\AA) & (\AA) & (s$^{-1}$) & & & (s$^{-1}$) & & \\
\hline
\endhead
 &  &  & 4s$^{2}$4p~$^{2}$P$^{\circ}_{1/2}$ & 4s$^{2}$6s~$^{2}$S$_{1/2}$ & 235.7225 & 0.0010 &  & 5.56e+09 & 0.64 & D+ &  &  &  \\
320 & 331.650 & 0.020 & 4s4p$^{2}$~$^{2}$D$_{3/2}$ & 4s4p($^{1}$P$^{\circ}$)5s~$^{2}$P$^{\circ}_{3/2}$ & 331.675 & 0.004 & -0.025 & 2.43e+09 & 0.22 & D+ & 6.03e+08 & P96 &  \\
6200* & 332.83 & 0.03 & 4s4p$^{2}$~$^{2}$D$_{5/2}$ & 4s4p($^{1}$P$^{\circ}$)5s~$^{2}$P$^{\circ}_{3/2}$ & 332.837 & 0.004 & -0.01 & 1.03e+10 & 0.15 & D+ & 2.45e+09 & P96 &  \\
6200* & 332.83 & 0.03 & 4s4p$^{2}$~$^{2}$D$_{3/2}$ & 4s4p($^{1}$P$^{\circ}$)5s~$^{2}$P$^{\circ}_{1/2}$ & 332.854 & 0.006 & -0.02 & 8.55e+09 & 0.26 & D+ & 2.01e+09 & P96 &  \\
160d & 615.066 & 0.016 & 4s4p$^{2}$~$^{4}$P$_{5/2}$ & 4p$^{3}$~$^{2}$D$^{\circ}_{5/2}$ & 615.075 & 0.005 & -0.009 & 2.75e+08 & 0.18 & D+ & 2.33e+08 & P96,TW &  \\
2100bl(Kr VII) & 617.191 & 0.016 & 4s4p$^{2}$~$^{2}$D$_{3/2}$ & 4p$^{3}$~$^{2}$P$^{\circ}_{1/2}$ & 617.206 & 0.006 & -0.016 & 1.09e+10 & 0.39 & C+ & 9.23e+09 & P96,TW & X \\
2200u & 617.368 & 0.016 & 4s$^{2}$4p~$^{2}$P$^{\circ}_{3/2}$ & 4s4p$^{2}$~$^{2}$S$_{1/2}$ & 617.377 & 0.004 & -0.008 & 1.42e+09 & 0.03 & E & 7.34e+08 & P96,TW & X \\
1100bl(Kr IV) & 622.941 & 0.008 & 4s4p$^{2}$~$^{4}$P$_{5/2}$ & 4p$^{3}$~$^{2}$D$^{\circ}_{3/2}$ & 622.935 & 0.004 & 0.005 & 6.10e+09 & 0.67 & D+ & 4.83e+09 & P96,TW & X \\
13p & 799.71 & 0.04 & 4s4p$^{2}$~$^{2}$P$_{1/2}$ & 4p$^{3}$~$^{2}$P$^{\circ}_{3/2}$ & 799.705 & 0.011 & 0.00 & 1.02e+07 & 0.00 & E & 8.52e+04 & P96 &  \\
110w & 810.647 & 0.016 & 4s4p$^{2}$~$^{2}$P$_{1/2}$ & 4p$^{3}$~$^{2}$P$^{\circ}_{1/2}$ & 810.656 & 0.009 & -0.009 & 3.19e+09 & 0.15 & C & 2.42e+09 & P96,TW &  \\
150w & 822.585 & 0.016 & 4s4p$^{2}$~$^{2}$P$_{3/2}$ & 4p$^{3}$~$^{2}$P$^{\circ}_{3/2}$ & 822.587 & 0.010 & -0.002 & 5.51e+09 & 0.16 & C+ & 4.02e+09 & P96,TW &  \\
33 & 998.961 & 0.016 & 4s$^{2}$5p~$^{2}$P$^{\circ}_{1/2}$ & 4s$^{2}$6s~$^{2}$S$_{1/2}$ & 998.937 & 0.012 & 0.024 & 3.16e+09 & 0.83 & C+ &  & TW* &  \\
54 & 1002.737 & 0.008 & 4s$^{2}$4p~$^{2}$P$^{\circ}_{3/2}$ & 4s4p$^{2}$~$^{4}$P$_{1/2}$ & 1002.735 & 0.007 & 0.001 & 9.05e+06 & 0.16 & D+ & 8.76e+06 & P96,TW &  \\
9 & 1011.138 & 0.016 & 4s$^{2}$5s~$^{2}$S$_{1/2}$ & 4s4p($^{3}$P$^{\circ}$)4d~$^{2}$P$^{\circ}_{3/2}$ & 1011.142 & 0.013 & -0.003 & 1.20e+08 & 0.38 & C & 9.44e+07 & P96,TW &  \\
37q,l & 1015.770 & 0.020 & 4s4p$^{2}$~$^{2}$P$_{1/2}$ & 4p$^{3}$~$^{4}$S$^{\circ}_{3/2}$ & 1015.753 & 0.010 & 0.017 & 1.77e+08 & 0.06 & D & 9.54e+07 & P96 &  \\
130w & 1024.878 & 0.016 & 4s$^{2}$5p~$^{2}$P$^{\circ}_{3/2}$ & 4s$^{2}$6s~$^{2}$S$_{1/2}$ & 1024.900 & 0.012 & -0.022 & 5.78e+09 & 0.82 & C+ &  & TW* &  \\
170w & 1950.20 & 0.04 & 4s$^{2}$5s~$^{2}$S$_{1/2}$ & 4s$^{2}$5p~$^{2}$P$^{\circ}_{3/2}$ & 1950.22 & 0.03 & -0.02 & 2.45e+09 & 0.65 & C+ & 2.32e+09 & P96 &  \\
80w & 2051.06 & 0.04 & 4s$^{2}$5s~$^{2}$S$_{1/2}$ & 4s$^{2}$5p~$^{2}$P$^{\circ}_{1/2}$ & 2051.03 & 0.03 & 0.03 & 1.06e+09 & 0.65 & C+ & 1.01e+09 & P96 &  \\
 & 12330 & 4 & 4s$^{2}$4p~$^{2}$P$^{\circ}_{1/2}$ & 4s$^{2}$4p~$^{2}$P$^{\circ}_{3/2}$ & 12328.6 & 1.5 & 1 &  &  &  &  & S17 &  \\
\hline
\vspace{-1cm}
\end{longtable}
\begin{hangparas}{.10in}{1}
\textbf{} \\
$^{a}$ Average relative observed line intensities in arbitrary units are expressed on a uniform scale corresponding to Boltzmann modeling with an effective excitation temperature of 7.6 eV for the Kr VI spectrogram observed by us (see Section \ref{sec:int_model}). The intensity value is followed by the line characters: bl---blended by other line (the blending spectrum is given in parentheses), d---diffused line, l---shaded to longer wavelengths, p---perturbed by close line, q---asymmetric line, u---unresolved line, w---wide line, *---intensity is shared by multiple transitions indicated. \\
$^{b}$ Observed and Ritz wavelengths are given in vacuum for wavenumbers $\sigma\ge 50000$ cm$^{-1}$, and in standard air for $\sigma < 50000$ cm$^{-1}$. The conversion between vacuum and air wavelengths was made with the help of the five-parameter formula given by \citep{Peck1972-index}. The quantity given in parentheses is the uncertainty in the last digit. The columns for observed wavenumber (in vacuum) and its uncertainty are omitted from this condensed table. \\
$^{c}$ Lower and upper level designations from Table \ref{tab:lev_KrVI}. \\
$^{d}$ Difference between the observed and Ritz wavelengths. \\
$^{e}$ Transition probability (gA-value) and absolute cancellation factor |CF| from the HFR-B model described in Section \ref{sec:theo_HFR}. \\
$^{f}$ Accuracy code of the gA-values are explained in Section \ref{sec:tp_eval}. \\
$^{g}$ gA-values from R16---\citet{Rauch_2016_Kr-VI}. \\
$^{h}$ Line references: P96---\citet{Pagan_1995_Kr6_exp, Pagan_1996_Kr6_os}, S17---\citet{Sterling_2017_Kr6_PNe}, TW---previous reported wavelength from P96 is replaced by this work, and those with an extra * denote newly observed lines. \\
$^{i}$ Comment: X---line excluded from the intensity averaging described in Section \ref{sec:int_model}. \\ \\
(This table is available in its entirety in machine-readable form as supplementary material.)
\end{hangparas}
\end{scriptsize}
\end{landscape}

%%%%%%%%%%%%%%%%%%%%%%%%%%%%%%%%%%%%%%%%%%
%%%%%%%%%%%%%% Table -- Levels %%%%%%%%
%%%%%%%%%%%%%%%%%%%%%%%%%%%%%%%%%%%%%%%%%%

\setlength\LTleft{10pt}
\setlength\LTright{0pt}
\begin{landscape}
\begin{scriptsize}
\begin{longtable}{lcccclclccl}
\caption{Optimized energy levels of Kr VI}
\label{tab:lev_KrVI}
\\
\hline
\hline
Level & Energy$^{a}$ & Unc.$^{b}$ & \multicolumn{5}{c}{Leading LS Compositions$^c$} & $\Delta$E$_{O-C}$$^{d}$ & No. of Lines$^{e}$ & Comm.$^{f}$ \\
\cline{4-8}
& (cm$^{-1}$) & (cm$^{-1}$) & P1 & P2 & Comp2 & P3 & Comp3 & (cm$^{-1}$) & \\
\hline
\endfirsthead
\multicolumn{9}{c}{Continuation of Table \ref{tab:lev_KrVI}}\\
\hline
\hline
Level & Energy$^{a}$ & Unc.$^{b}$ & \multicolumn{5}{c}{Leading LS Compositions$^c$} & $\Delta$E$_{O-C}$$^{d}$ & No. of Lines$^{e}$ & Comm.$^{f}$ \\
\cline{4-8}
& (cm$^{-1}$) & (cm$^{-1}$) & P1 & P2 & Comp2 & P3 & Comp3 & (cm$^{-1}$) & \\
\hline
\endhead
4s$^{2}$4p~$^{2}$P$^{\circ}_{1/2}$ & 0.0 & 0.8 & 98 & & ... & & ... & -20 & 9 & \\
4s$^{2}$4p~$^{2}$P$^{\circ}_{3/2}$ & 8109.0 & 0.7 & 98 & & ... & & ... & 18 & 12 & \\
4s4p$^{2}$~$^{4}$P$_{1/2}$ & 107836.2 & & 99 & & ... & & ... & 23 & 9 & B \\
4s4p$^{2}$~$^{4}$P$_{3/2}$ & 111193.2 & 0.9 & 99 & & ... & & ... & -18 & 14 & \\
4s4p$^{2}$~$^{4}$P$_{5/2}$ & 115481.7 & 1.2 & 98 & 2 & 4s4p$^{2}$($^{1}$D)~$^{2}$D & & ... & -7 & 15 & \\
4s4p$^{2}$~$^{2}$D$_{3/2}$ & 141675.1 & 1.3 & 89 & 9 & 4d~$^{2}$D & & ... & -78 & 20 & \\
4s4p$^{2}$~$^{2}$D$_{5/2}$ & 142728.0 & 1.4 & 88 & 8 & 4d~$^{2}$D & & ... & 79 & 22 & \\
4s4p$^{2}$~$^{2}$S$_{1/2}$ & 170084.7 & 1.1 & 74 & 23 & 4s4p$^{2}$($^{3}$P)~$^{2}$P & & ... & 10 & 15 & \\
4s4p$^{2}$~$^{2}$P$_{1/2}$ & 180338.6 & 1.1 & 74 & 24 & 4s4p$^{2}$($^{1}$S)~$^{2}$S & & ... & -32 & 14 & \\
4s4p$^{2}$~$^{2}$P$_{3/2}$ & 183817.0 & 1.2 & 97 & & ... & & ... & 23 & 22 & \\
4s$^{2}$4d~$^{2}$D$_{3/2}$ & 222123.3 & 1.3 & 89 & 9 & 4s4p$^{2}$($^{1}$D)~$^{2}$D & & ... & 7 & 16 & \\
4s$^{2}$4d~$^{2}$D$_{5/2}$ & 223041.6 & 1.2 & 89 & 8 & 4s4p$^{2}$($^{1}$D)~$^{2}$D & & ... & -7 & 13 & \\
4s$^{2}$5s~$^{2}$S$_{1/2}$ & 275381.0 & 1.3 & 98 & & ... & & ... & 0 & 10 & \\
4p$^{3}$~$^{2}$D$^{\circ}_{3/2}$ & 276012.0 & 1.0 & 60 & 16 & 4s4p($^{3}$P$^{\circ}$)4d~$^{2}$D$^{\circ}$ & 16 & 4p$^{3}$~$^{4}$S$^{\circ}$ & -32 & 7 & \\
4p$^{3}$~$^{2}$D$^{\circ}_{5/2}$ & 278063.6 & 1.2 & 78 & 20 & 4s4p($^{3}$P$^{\circ}$)4d~$^{2}$D$^{\circ}$ & & ... & 3 & 6 & \\
4p$^{3}$~$^{4}$S$^{\circ}_{3/2}$ & 278787.7 & 1.1 & 80 & 14 & 4p$^{3}$~$^{2}$D$^{\circ}$ & & ... & 2 & 8 & \\
4p$^{3}$~$^{2}$P$^{\circ}_{1/2}$ & 303695.5 & 1.5 & 84 & 10 & 4s4p($^{3}$P$^{\circ}$)4d~$^{2}$P$^{\circ}$ & & ... & -72 & 3 & \\
4p$^{3}$~$^{2}$P$^{\circ}_{3/2}$ & 305384.7 & 1.7 & 76 & 11 & 4s4p($^{3}$P$^{\circ}$)4d~$^{2}$P$^{\circ}$ & & ... & 52 & 5 & \\
4s4p($^{3}$P$^{\circ}$)4d~$^{4}$F$^{\circ}_{3/2}$ & 314778 & 3 & 98 & & ... & & ... & -212 & 3 & N \\
4s4p($^{3}$P$^{\circ}$)4d~$^{4}$F$^{\circ}_{5/2}$ & 316500.5 & 1.6 & 97 & & ... & & ... & -139 & 4 & N \\
4s4p($^{3}$P$^{\circ}$)4d~$^{4}$F$^{\circ}_{7/2}$ & 319460 & 4 & 97 & & ... & & ... & 344 & 1 & N \\
4s4p($^{3}$P$^{\circ}$)4d~$^{4}$F$^{\circ}_{9/2}$ & 322700 & 400 & 99 & & ... & & ... & & & L \\
4s$^{2}$5p~$^{2}$P$^{\circ}_{1/2}$ & 324121.3 & 1.3 & 96 & & ... & & ... & -15 & 4 & \\
4s$^{2}$5p~$^{2}$P$^{\circ}_{3/2}$ & 326657.2 & 1.2 & 96 & & ... & & ... & 16 & 8 & \\
4s4p($^{3}$P$^{\circ}$)4d~$^{4}$P$^{\circ}_{5/2}$ & 331957.8 & 1.6 & 67 & 26 & 4s4p($^{3}$P$^{\circ}$)4d~$^{4}$D$^{\circ}$ & & ... & -74 & 6 & \\
4s4p($^{3}$P$^{\circ}$)4d~$^{4}$D$^{\circ}_{3/2}$ & 333129 & 3 & 56 & 41 & 4s4p($^{3}$P$^{\circ}$)4d~$^{4}$P$^{\circ}$ & & ... & -87 & 4 & \\
4s4p($^{3}$P$^{\circ}$)4d~$^{4}$D$^{\circ}_{1/2}$ & 333940 & 3 & 84 & 14 & 4s4p($^{3}$P$^{\circ}$)4d~$^{4}$P$^{\circ}$ & & ... & -75 & 2 & \\
4s4p($^{3}$P$^{\circ}$)4d~$^{4}$P$^{\circ}_{1/2}$ & 338027 & 4 & 85 & 14 & 4s4p($^{3}$P$^{\circ}$)4d~$^{4}$D$^{\circ}$ & & ... & -49 & 1 & \\
4s4p($^{3}$P$^{\circ}$)4d~$^{4}$D$^{\circ}_{7/2}$ & 338120.8 & 2.1 & 97 & 2 & 4s4p($^{3}$P$^{\circ}$)4d~$^{4}$F$^{\circ}$ & & ... & 97 & 3 & \\
4s4p($^{3}$P$^{\circ}$)4d~$^{4}$P$^{\circ}_{3/2}$ & 338366.6 & 2.3 & 56 & 43 & 4s4p($^{3}$P$^{\circ}$)4d~$^{4}$D$^{\circ}$ & & ... & 12 & 4 & \\
4s4p($^{3}$P$^{\circ}$)4d~$^{4}$D$^{\circ}_{5/2}$ & 338450.2 & 2.0 & 72 & 25 & 4s4p($^{3}$P$^{\circ}$)4d~$^{4}$P$^{\circ}$ & & ... & 56 & 3 & \\
4s4p($^{3}$P$^{\circ}$)4d~$^{2}$D$^{\circ}_{3/2}$ & 343188.7 & 1.9 & 63 & 19 & 4s4p($^{1}$P$^{\circ}$)4d~$^{2}$D$^{\circ}$ & 13 & 4p$^{3}$~$^{2}$D$^{\circ}$ & -28 & 6 & \\
4s4p($^{3}$P$^{\circ}$)4d~$^{2}$D$^{\circ}_{5/2}$ & 343505.9 & 1.4 & 58 & 18 & 4s4p($^{1}$P$^{\circ}$)4d~$^{2}$D$^{\circ}$ & 12 & 4p$^{3}$~$^{2}$D$^{\circ}$ & -20 & 7 & \\
4s4p($^{3}$P$^{\circ}$)4d~$^{2}$F$^{\circ}_{5/2}$ & 352545.5 & 2.3 & 62 & 30 & 4s4p($^{1}$P$^{\circ}$)4d~$^{2}$F$^{\circ}$ & & ... & -520 & 3 & \\
4s4p($^{3}$P$^{\circ}$)4d~$^{2}$F$^{\circ}_{7/2}$ & 359036 & 3 & 50 & 28 & 4f~$^{2}$F$^{\circ}$ & 20 & 4s4p($^{3}$P$^{\circ}$)4d~$^{2}$F$^{\circ}$ & 489 & 3 & \\
4s$^{2}$4f~$^{2}$F$^{\circ}_{5/2}$ & 360724 & 3 & 57 & 22 & 4s4p($^{3}$P$^{\circ}$)4d~$^{2}$F$^{\circ}$ & 19 & 4s4p($^{1}$P$^{\circ}$)4d~$^{2}$F$^{\circ}$ & 34 & 3 & N \\
4s$^{2}$4f~$^{2}$F$^{\circ}_{7/2}$ & 360843 & 3 & 30 & 67 & 4s4p($^{3}$P$^{\circ}$)4d~$^{2}$F$^{\circ}$ & & ... & -146 & 2 & N \\
4s4p($^{3}$P$^{\circ}$)4d~$^{2}$P$^{\circ}_{3/2}$ & 374279.1 & 1.7 & 80 & 12 & 4p$^{3}$~$^{2}$P$^{\circ}$ & 5 & 4s4p($^{1}$P$^{\circ}$)4d~$^{2}$D$^{\circ}$ & 241 & 4 & \\
4s4p($^{3}$P$^{\circ}$)4d~$^{2}$P$^{\circ}_{1/2}$ & 377252 & 3 & 87 & 10 & 4p$^{3}$~$^{2}$P$^{\circ}$ & & ... & -54 & 3 & \\
4s4p($^{1}$P$^{\circ}$)4d~$^{2}$D$^{\circ}_{3/2}$ & 390597 & 4 & 68 & 11 & 4s4p($^{3}$P$^{\circ}$)4d~$^{2}$D$^{\circ}$ & 7 & 4p$^{3}$~$^{2}$D$^{\circ}$ & 46 & 3 & \\
4s4p($^{1}$P$^{\circ}$)4d~$^{2}$P$^{\circ}_{1/2}$ & 391600 & 4 & 84 & 9 & 4s4p($^{3}$P$^{\circ}$)5s~$^{4}$P$^{\circ}$ & & ... & 29 & 3 & N \\
4s4p($^{1}$P$^{\circ}$)4d~$^{2}$D$^{\circ}_{5/2}$ & 391877 & 3 & 74 & 14 & 4s4p($^{3}$P$^{\circ}$)4d~$^{2}$D$^{\circ}$ & 8 & 4p$^{3}$~$^{2}$D$^{\circ}$ & 157 & 2 & \\
4s4p($^{3}$P$^{\circ}$)5s~$^{4}$P$^{\circ}_{1/2}$ & 392099.6 & 2.1 & 88 & 9 & 4s4p($^{1}$P$^{\circ}$)4d~$^{2}$P$^{\circ}$ & & ... & 126 & 4 & N \\
4s4p($^{1}$P$^{\circ}$)4d~$^{2}$P$^{\circ}_{3/2}$ & 393014.8 & 2.4 & 84 & 4 & 4s4p($^{1}$P$^{\circ}$)4d~$^{2}$D$^{\circ}$ & & ... & 195 & 6 & \\
4s4p($^{3}$P$^{\circ}$)5s~$^{4}$P$^{\circ}_{3/2}$ & 394815 & 3 & 93 & 3 & 4s4p($^{3}$P$^{\circ}$)5s~$^{2}$P$^{\circ}$ & & ... & 383 & 5 & \\
4s4p($^{1}$P$^{\circ}$)4d~$^{2}$F$^{\circ}_{7/2}$ & 398675.0 & 2.5 & 45 & 40 & 4f~$^{2}$F$^{\circ}$ & 11 & 4s4p($^{3}$P$^{\circ}$)4d~$^{2}$F$^{\circ}$ & -152 & 2 & \\
4s4p($^{1}$P$^{\circ}$)4d~$^{2}$F$^{\circ}_{5/2}$ & 399592 & 3 & 45 & 38 & 4f~$^{2}$F$^{\circ}$ & 12 & 4s4p($^{3}$P$^{\circ}$)4d~$^{2}$F$^{\circ}$ & -108 & 4 & \\
4s4p($^{3}$P$^{\circ}$)5s~$^{4}$P$^{\circ}_{5/2}$ & 399613 & 10 & 99 & & ... & & ... & -256 & 3 & \\
4s4p($^{3}$P$^{\circ}$)5s~$^{2}$P$^{\circ}_{1/2}$ & 403438 & 3 & 96 & & ... & & ... & 329 & 4 & \\
4s4p($^{3}$P$^{\circ}$)5s~$^{2}$P$^{\circ}_{3/2}$ & 408514 & 3 & 93 & 3 & 4s4p($^{3}$P$^{\circ}$)5s~$^{4}$P$^{\circ}$ & & ... & -230 & 6 & \\
4s$^{2}$6s~$^{2}$S$_{1/2}$ & 424227.7 & 1.6 & 98 & 2 & 4p$^{2}$($^{1}$S)6s~$^{2}$S & & ... & 0 & 2 & N \\
4s4p($^{1}$P$^{\circ}$)5s~$^{2}$P$^{\circ}_{1/2}$ & 442107 & 5 & 71 & 26 & 6p~$^{2}$P$^{\circ}$ & & ... & -247 & 4 & \\
4s4p($^{1}$P$^{\circ}$)5s~$^{2}$P$^{\circ}_{3/2}$ & 443175 & 4 & 74 & 20 & 6p~$^{2}$P$^{\circ}$ & & ... & -205 & 6 & \\
\hline
\end{longtable}
\vspace{-10pt}
\begin{hangparas}{0.10in}{1}
\textbf{} \\
$^{a}$ Optimized energy values derived using LOPT code (see Section \ref{sec:lev_opt}).\\
$^{b}$ Uncertainties resulting from the level optimization procedure for all observed levels. They are given on the level of 1-SD. They correspond to uncertainties of level separations from the base level 4s4p$^{2}$~$^{4}$P$_{1/2}$. The given values of their uncertainties should be combined in quadrature to derive the absolute uncertainties for any excited level with respect to the ground level. \\
$^{c}$ The LS coupling percentage composition vectors from our HFR-B model, wherein P1 refers to the first percentage value for the configuration and term given in the first column of the table. The remaining percentage (P2, P3) values are provided with their corresponding LS components (see Section \ref{sec:theo_HFR}). \\
$^{d}$ Differences between observed and calculated energies from the parametric least-squares fitting (LSF). Blank for unobserved levels. \\
$^{e}$ Number of lines involved in the LOPT scheme (see Section \ref{sec:lev_opt}). \\
$^{f}$ Comment: B---the given level was selected as the base level for deriving the level uncertainties of all other known levels, L---the given theoretical level value and its uncertainty are from the LSF of the HFR-B model, N---new level established by us (see Section \ref{sec:new_levels}).
\end{hangparas}
\end{scriptsize}
\end{landscape}
%%%%%%%%%%%%%%%%%%%%%%%%%%%%%%%%%%%%%%%%%%%%%%%%%%%%%%%%%
%%%%%%%%%%%%%% Table 2 (Levels) End %%%%%%%%%%
%%%%%%%%%%%%%%%%%%%%%%%%%%%%%%%%%%%%%%%%%%%%%%%%%%%%%%%%%

\subsection{Analysis of the spectrum} \label{sec:analysis_spec}

Our analysis of the Kr VI spectrum began with the search for its previously reported lines \citep{Pagan_1995_Kr6_exp, Pagan_1996_Kr6_os, Farias_2011_Kr-VI} in our global line list. This line list contains about 2700 spectral features, of varying species and ionization states, with their calibrated wavelengths or wavenumbers, intensities, and line characteristics. The impurity and second-order diffraction lines were marked and excluded from the line list. We confirmed most of the observations made by Pagan et al. \citep{Pagan_1995_Kr6_exp, Pagan_1996_Kr6_os}; however, the Kr VI's identification by \citet{Farias_2011_Kr-VI} could not be verified (see details in Section \ref{sec:measur_F11}). Nonetheless, we provided detailed descriptions of these previous experiments with thorough discussions of the present analyses in the sub-sections below, and explanations for our newly established levels of Kr VI are given in Section \ref{sec:new_levels}. \\

\subsubsection {Measurements of Pagan et al.} \label{sec:measur_P96}

The spectrograms used by Pagan et al. \citep{Pagan_1995_Kr6_exp, Pagan_1996_Kr6_os} for the analysis of Kr VI were recorded on two different 3~m NIVS of the same grating grooves density 1200 lines mm$^{-1}$, but blazed at two different wavelengths (1380~\AA~and 1200~\AA). Both instruments had a linear reciprocal dispersion of $\approx$ 2.77 \AA~mm$^{-1}$. In one of the instruments, a theta-pinch source built at the Lund Institute of Technology \citep{Trigueiros_1988_Kr6, Pettersson_1982_O3_theta-pinch} was used to record the spectrum for the 432--2540~\AA~region. In other experiments, the second instrument at the Centro de Investigaciones Opticas, Argentina, operated with a CD source \citep{Gallardo_1989_Light_source}, supplied the spectrum in the wavelength 233--456~\AA~range. The exposures were taken either on Kodak SWR or Ilford Q-2 plates. For spectrograms with theta-pinch, the relative positions of spectral lines were measured on the Zeiss Abbe with a photoelectric readout to visualize the spectral line profiles on an oscilloscope screen. A dedicated Grant comparator was used to measure the spectrograms with a CD source. The wavelength reduction was carried out with the help of internal standards provided by impurities of C I-II, O I, N I-II, and Si II lines, whose reference wavelengths were taken from \citet{Kaufman_1974_Ref_WL_15to25}. A total of 142 classified lines were reported, of which 108 and 33 were from the theta-pinch and CD sources, respectively. The reported wavelengths from the theta-pinch source were averaged from more than four exposures, representing different experimental conditions. The reported intensities of lines were based on the visual estimates on the oscilloscope screen. The measurement uncertainties were 0.01~\AA~and 0.005~\AA, for unperturbed lines of the first- and second-diffraction orders, respectively. The identified lines belong to the following transition arrays: 4s$^{2}$4p--\{4s4p$^{2}$, 4s$^{2}$(4d+5s)\}, 4s4p$^{2}$--\{4p$^{3}$, 4s$^2$5p, 4s4p(4d+5s)\}, 4s$^{2}$4d--(4p$^3$, 4s$^{2}$5p, 4s4p4d), and 4s$^2$5s--\{4s$^2$5p, 4s4p(4d+5s)\}. As a result, energy values for 44 known levels were reported. \\

It should be noted that Pagan et al. \citep{Pagan_1995_Kr6_exp, Pagan_1996_Kr6_os} observations are the primary data source for Kr VI data compiled by \citet{Saloman_2007_Kr_compil}, and accordingly for its accessible data in the NIST ASD \citep{nist}. Therefore, we intended to verify these existing identifications at first. In our spectrograms, all spectral lines except a few weak lines and/or lines of highly excited levels, for example, those of 4s4p4d and 4s4p5s configurations, which fall in the shorter wavelength region ($\lambda<400$~\AA), reported by Pagan et al. \citep{Pagan_1995_Kr6_exp, Pagan_1996_Kr6_os} were appeared with certainty; thus, we confirmed the observation made by Pagan et al. However, our spectrograms were better in terms of the reciprocal dispersion, at least by a factor of two smaller, and as a result of it, the lines could be better resolved on our plates, and an improved wavelength accuracy was expected for our wavelength measurements. {For example, the 4s$^{2}$4p~$^{2}$P$^{\circ}_{3/2}$--4s$^{2}$5s~$^{2}$S$_{1/2}$ transition was not observed by Pagan et al. in their spectrograms, possibly due to the strong blending of O III lines at 374.16~\AA~\citep{nist}. Instead, its roughly measured wavelength was quoted from the \citet{Tauheed_1990_Kr-VI}. This Kr VI line is resolved from its close-by O III line in our spectrogram, and was observed at 374.151~\AA~as one of the strongest lines.} Several additional lines, previously not reported or observed by Pagan et al., were found on our plates. To support and confirm all these observations, an additional Boltzmann plot analysis, described in Section \ref{sec:int_model}, was performed for them. \\

\subsubsection {Measurements of Farias et al.} \label{sec:measur_F11} 

The spectrograms used by \citet{Farias_2011_Kr-VI} -- which extended the analysis for the levels of Kr VI 4s4p5p, 4p$^2$4d, and 4s4p4f -- were recorded on the same instruments with the same light sources as discussed in section \ref{sec:measur_P96}. However, these spectrograms fall in the wavelength region of 240–2600 {\AA}. The exposures were taken on the Ilford Q-2 plates. The line intensities were determined using the visual estimates of plate blackening. The claimed wavelength uncertainty of the measurement was 0.01~\AA. A total of 87 lines were observed, which belong to the transitions arrays:- \{4p$^{3}$, 4s$^{2}$5p, 4s4p(4d+5s)\}--\{4s4p(5p+4f), 4p$^2$4d\}. Consequently, 17 levels (out of a total of 18) of 4s4p5p with a few supporting levels of 4s4p4f (2 levels) and 4p$^2$4d (6 levels) configurations were experimentally known. The levels reported by Farias et al. were mostly supported by multiple transitions:- 7 levels with 5 and 4 transitions, four levels with 3, 5 with 2, and 2 with a single transition. On the one hand, \citet{Farias_2011_Kr-VI} claimed that their findings were supported by the HFR calculations with the LSF adjustments, and the improved gA-values (after the LSF adjustments) helped them to identify new transitions. On the other hand, they did not provide any gA-values to substantiate the observed lines' intensities in their line list (see Tables 1 \& 4 of ref.~\citep{Farias_2011_Kr-VI}). Thus, it was difficult to verify their classification, for example, by correlating their observed lines' intensities with the theoretical gA-values. However, we were constrained to do it by taking gA-values from the published work of \citet{Rauch_2016_Kr-VI} or our HFR-B calculations with the additional inclusion of levels from Farias et al. into the LSF. The obtained correlation coefficient for the branching fraction comparison, between theoretical (from gA-values) and observed (from original lines' intensities) for unperturbed lines, was smaller than that of Pagan et al. \citep{Pagan_1995_Kr6_exp, Pagan_1996_Kr6_os} observations. In fact the wavelength dependent response of the instrument, including that of the registration equipment, is not a constant for the entire wavelength range; in reality, it drastically falls for the blazed grating instruments to the either side of the blazed wavelength; thus, these comparisons are limited and could be accurate within a short span of wavelengths range. An improved comparison for a wider wavelength region could be achieved with modeled or corrected line intensities from Boltzmann plot analysis (see Section \ref{sec:int_model} for details). However, a separate Boltzmann modeling for Farias et al.'s observations, with 87 transitions and their 25 upper energy levels, became futile without including Pagan et al.'s observations into the model. Farias et al. and Pagan et al. used the same recording instrument and light sources for their study; it is unclear whether their spectrograms are the same or different. Apart from this, Farias et al. did not describe the appearance of Pagan et al. observed transitions in the spectrogram; nonetheless, they mentioned that two classified transitions (lines at $\lambda$ = 475.62 and 1045.22~\AA~in Table 4 of ref. \citep{Farias_2011_Kr-VI}) were already reported by Pagan et al. with the same intensities and line character. With this indicator, we searched and checked for their lines in the combined line list of Kr ions published by this group, for which data were accessed from refs. \citep{Saloman_2007_Kr_compil, nist}. Surprisingly, we found five additional perfect matches (the same wavelength with the similar intensity and line character):- two Kr VI lines (910.47 and 2051.06~\AA~(in air), in the line list of Pagan et al. \citep{Pagan_1995_Kr6_exp, Pagan_1996_Kr6_os}) and and three Kr V lines (627.40, 628.83, and 1498.28~\AA, reported by \citet{Raineri_2002_Kr5}). {Therefore, we speculate that all the above observations reported by this group could be from the same (set of) experiments or spectrograms, as the case may be.} It should be noted that Farias et al.'s obtained levels are the highest known energy levels in Kr VI (see Figure \ref{fig:energy_lev}); thus, substantiation of Pagan et al.'s observations in the spectrum is a prerequisite for confirmation of the observation. By all these considerations, a combined Boltzmann plot analysis was performed for these observations \citep{Pagan_1995_Kr6_exp, Pagan_1996_Kr6_os, Farias_2011_Kr-VI} in the Kr VI spectrum. The required gA-values for this analysis were taken either from \citet{Rauch_2016_Kr-VI} published results or from our HFR-B calculations with the provisional incorporation of levels of Farias et al. into LSF (see Section \ref{sec:theo_HFR}). The initial excitation temperature of the source for this combined model was found to be $-59$ eV and $\approx6.5$ eV for a model with the observations of Pagan et al. itself. The corrected lines' intensities were derived by removing the wavelength dependency of the instrument and the non-linear response of photographic plates. The final effective excitation temperature of the source from these corrected intensities for the combined model was 24 eV; meanwhile, that for the model with the observations of Pagan et al. was about 8.1 eV (see Section \ref{sec:int_model} for further details). The correlation between theoretical (from gA-values) and observed (from corrected intensities) branching fractions was expected to be improved; however, no such improvement was found for the observations of \citet{Farias_2011_Kr-VI}. Also, several lines reported by Farias et al. significantly deviated, stronger than estimated, from their calculated intensities. {In addition to all these, our simple comparison of the theoretical spectrum with those observed by Farias et al. reveals the major pitfall of this observation.} \\

\begin{figure*}[ht!]
\centering
\advance\leftskip-2cm
\advance\rightskip-2cm
\includegraphics[scale=0.70]{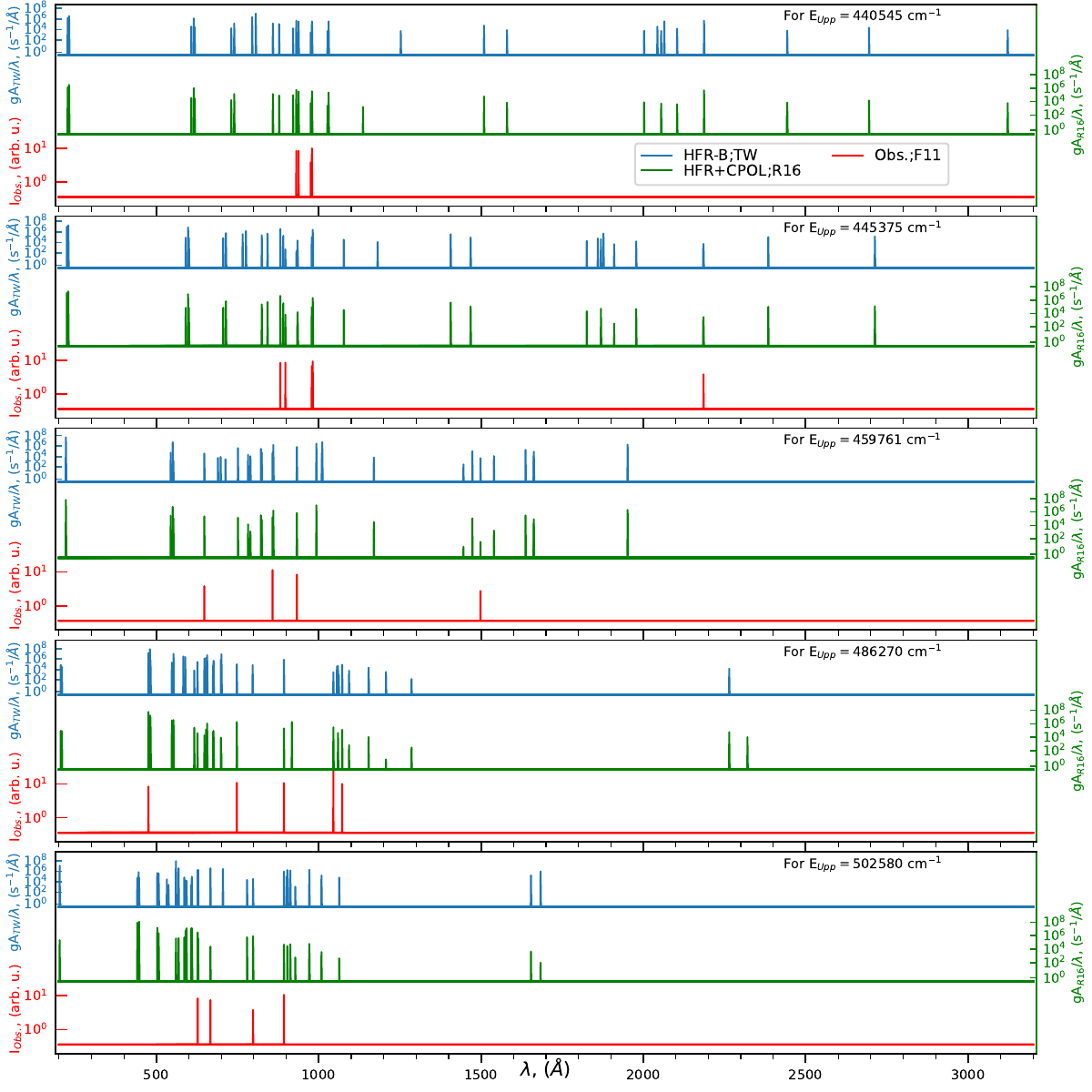}
\vspace{-15pt}
\caption{Comparisons of theoretical spectra, using transition data sets from the ad-hoc HFR-B model of this work (TW) and those of HFR+CPOL by R16--\citet{Rauch_2016_Kr-VI}, with the observed ones by \citet{Farias_2011_Kr-VI} for transitions originating from a group of selected upper energy levels (see text).} \label{fig:F11_spectra}
\end{figure*}

{As briefly described above, \citet{Rauch_2016_Kr-VI} included the levels established by Farias et al. in their theoretical HFR+CPOL model. Accordingly, they provided the radiative transition parameters for several possible lines between the known energy levels, including those of \citet{Farias_2011_Kr-VI}. It was noticed that the original level designation given by Farias et al. for several of their established energy levels differs from those of Rauch et al., albeit the levels involved have significant configuration interactions. In order to inspect this data set, an ad-hoc LSF calculation was made by incorporating Farias et al. energy levels in the HFR-B model described in Section \ref{sec:theo_HFR}. The levels were incorporated in this LSF according to their original designations by Farias et al., and radiative transition parameters, such as wavelengths, gA-values, and absolute CF-values, were generated for several possible lines between the levels involved. The ad-hoc HFR-B model and HFR+CPOL data sets representing Farias et al. observations are omitted from the line list for brevity. Both of these theoretical data sets (gA-values) agree reasonably well for all matching levels with the same level designations in Farias et al. and Rauch et al., and there are considerable disagreements for levels with differing designations. On the other hand, their comparison with Farias et al.'s observations is weakly correlated with these theoretical predictions. In Figure \ref{fig:F11_spectra}, we compare a few sample spectra synthesized using theoretical gA-values of HFR-B and HFR+CPOL data sets with those observed by Farias et al. for selected upper energy levels. This figure also shows that Farias et al. failed to observe several theoretically predicted strong lines, which fall in the recorded spectral region by these authors. In contrast, these authors observed and identified several theoretically weak lines, and their established energy levels were primarily based on such weak lines.} \\

Unlike the Pagan et al. reported transitions, only 30 lines reported by \citet{Farias_2011_Kr-VI} appeared on our plates, and only a limited number of levels were supported by multiple combinations of transitions. Moreover, their ionization states were ambiguous for several such lines on our plates. Nonetheless, we investigated them further with the help of a Boltzmann plot analysis:- a combined model showed initial and final temperatures to be 31 eV and 16 eV, respectively. Meanwhile, the same for the observed lines of Pagan et al. on our plates were 7.4 eV and 8.1 eV, respectively. These significant and abnormal differences between initial and final temperatures may be considered as an indicator for erroneous lines' identifications -- several weak lines appeared or were classified as strong lines. Due to these discrepancies and shortcomings, the observations of \citet{Farias_2011_Kr-VI} are omitted by us from the present classified list of lines and known levels of Kr VI, i.e., from Tables \ref{tab:lin_KrVI} \& \ref{tab:lev_KrVI}. Therefore, as we pointed out above, additional observations are needed to confirm these measurements.

\subsubsection{New levels of 4s4p(4d+5s) and \texorpdfstring{4s$^{2}$(4f+6s)}{} configurations} \label{sec:new_levels}

The observations of Pagan et al. \citep{Pagan_1995_Kr6_exp, Pagan_1996_Kr6_os} include only 18 known levels (out of 23) of 4s4p4d configuration, and the $^{4}$P$^{\circ}_{1/2}$ level of 4s4p5s was also unreported by them. The missing levels of 4s4p4d were:- four levels of 4s4p($^{3}$P$^{\circ}$)4d $^{4}$F$^{\circ}$ term and one $^{2}$P$^{\circ}_{1/2}$ of 4s4p($^{1}$P$^{\circ}$)4d. Since several additional lines, previously not observed by Pagan et al., were found in our spectrograms, we searched for these unknown levels. Our search process was assisted by semi-empirically produced radiative transition parameters, such as wavelengths, gA-values, and branching fractions, from an extensive HFR-B calculations with parametric LSF adjustments made with the help of Pagan et al. observed levels (see Section \ref{sec:theo_HFR}). Apart from these, we noticed that the levels of 4s$^2$4f were strongly admixed with the levels of 4s4p4d; thus, transitions from the levels of 4s$^2$4f could be expected on the same spectrum. \\

\begin{figure*}[ht!]
\centering
\advance\leftskip-2cm
\advance\rightskip-2cm
\includegraphics[scale=0.55]{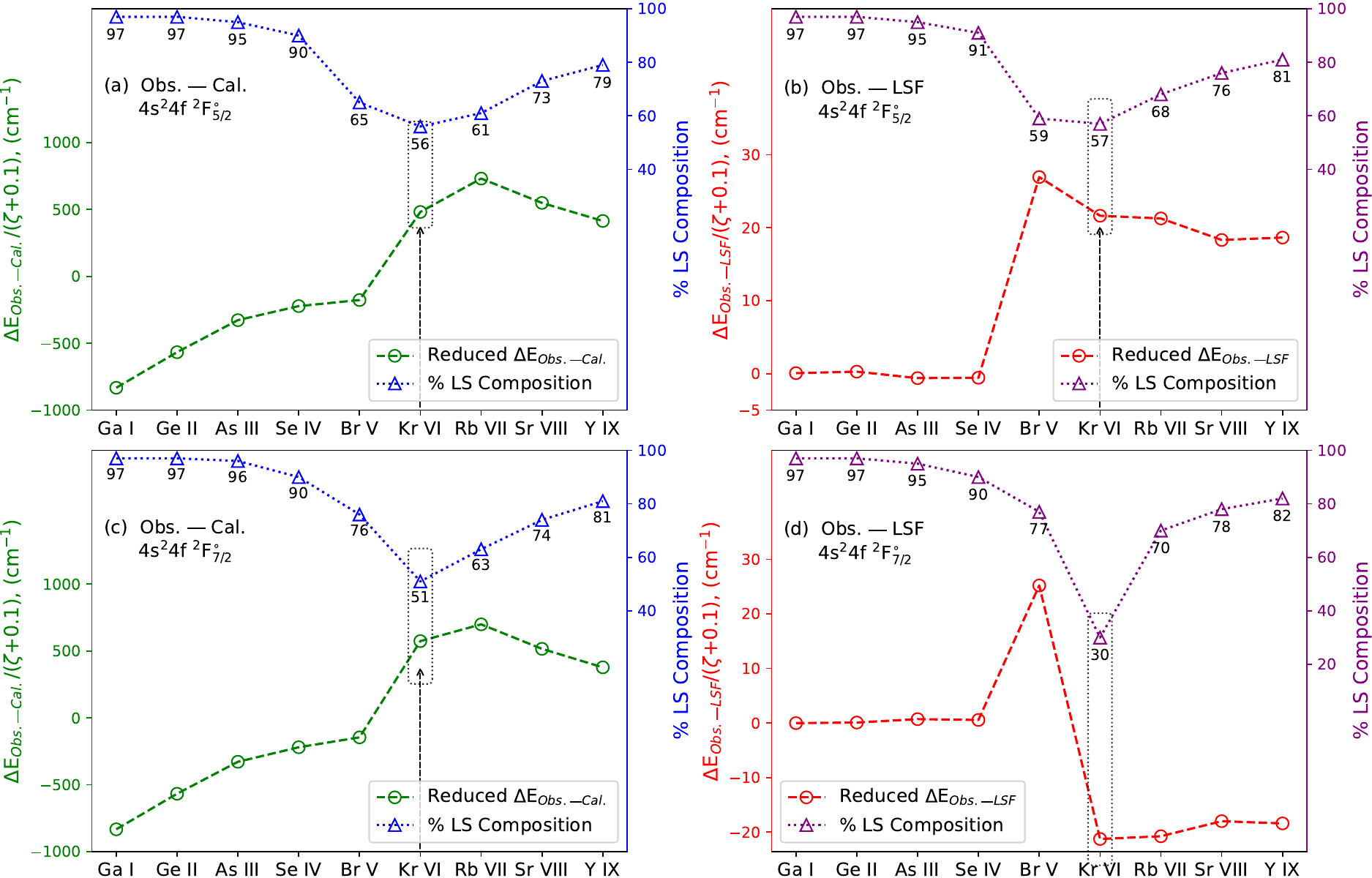}
\caption{Reduced energy differences between (a,c) observed and calculated values (i.e., ab initio energies from the HFR-B model) and (b,d) observed and LSF values from the HFR-B model for the 4s$^2$4f~$^{2}$F$^{\circ}_{5/2,~7/2}$ levels together with their respective \% LS compositions (from ab-initio and LSF calculations) along Ga I isoelectronic sequence (see text).}
\label{fig:4f_iso}
\end{figure*}

In the present work, we established the J = 3/2, 5/2, and 7/2 levels of 4s4p($^{3}$P$^{\circ}$)4d $^{4}$F$^{\circ}$ term at 314778 cm$^{-1}$, 316500.5 cm$^{-1}$, and 319460 cm$^{-1}$, respectively. The level with J = 3/2, 5/2 was supported by three and four transitions, respectively, whereas the J = 7/2 was solely based on a line at 490.248~\AA. The 4s4p($^{3}$P$^{\circ}$)4d~$^{4}$F$^{\circ}_{9/2}$ level is not making any allowed electric-dipole transitions to the lower levels; therefore, it is still unknown. Nevertheless, our LSF predicted this level at 322700 cm$^{-1}$ with an uncertainty of 400 cm$^{-1}$. The two missing levels 4s4p($^{1}$P$^{\circ}$)4d~$^{2}$P$^{\circ}_{1/2}$
and 4s4p($^{3}$P$^{\circ}$)5s $^{4}$P$^{\circ}_{1/2}$ were established at 391600 cm$^{-1}$ and 392099.6 cm$^{-1}$, respectively, with the help of four supporting transitions. All these transitions were included in the line list of Kr VI (see Table \ref{tab:lin_KrVI}). The $^{2}$F$^{\circ}$ levels 4s$^2$4f were previously reported by \citet{Tauheed_1990_Kr-VI}; however, those were subsequently rejected by Pagan et al. \citep{Pagan_1995_Kr6_exp}. We established J = 5/2 and 7/2 levels of 4s$^2$4f~$^{2}$F$^{\circ}$ term at 360724 cm$^{-1}$ (with three supporting lines) and 360843 cm$^{-1}$ (with two lines), respectively. These values are close to their predicted energies from LSF by Pagan et al. \citep{Pagan_1995_Kr6_exp} and from our LSF for the HFR-B model. As described above, the levels of 4s$^2$4f have significant interaction with those of 4s4p4d along the sequence~\citep{Pagan_1995_Kr6_exp, Litzen_1989_Ga_iso_Rb_Y}. Figure \ref{fig:4f_iso} was made to support and illustrate our findings, in which two isoelectronic plots -- based on our computed ab-initio and LSF energies from the HFR-B model (see section \ref{sec:theo_HFR}) -- were plotted for the levels of 4s$^2$4f together with their corresponding percentage (\%) LS composition vectors. The required experimental energy level data of Ga-like ions (Ga I through Y IX), for making these plots for 4s$^2$4f~$^{2}$F$^{\circ}_{5/2,~7/2}$ levels and also for performing their respective LSF calculations with HFR-B models, were taken from the NIST ASD \citep{nist} wherever applicable, otherwise from the quoted references: As III~ \citep{Churilov_1996_As3_As4}, Br V~\citep{Riyaz_2014_Br5}, Y IX \citep{Litzen_1989_Ga_iso_Rb_Y}. These plots also describe the variation of the \% LS composition vectors of the 4s$^2$4f~$^{2}$F$^{\circ}$ levels along the sequence from ab-initio and LSF calculations with the HFR-B model. Although configuration interactions between $^{2}$F$^{\circ}$ levels of the both 4s$^2$4f and 4s4p($^{1,3}$P$^{\circ}$)4d are stronger in Kr VI, but those for J = 7/2 levels are more significant. Thus, J = 7/2 levels are more admixed with each other and their LS purities are lower than that of J = 5/2 levels (see Figure \ref{fig:4f_iso} and/or refer to Table \ref{tab:lev_KrVI} for their LS purity and composition vectors). Last not but the least, the 4s$^2$6s~$^2$S$_{1/2}$ was established at 424227.7 cm$^{-1}$ due to two lines terminating to 4s$^2$5p~$^{2}$P$^{\circ}_J$ levels at 998.961 \AA~(for J = 1/2) and 1024.878 \AA~(for J = 3/2). \\

\subsection{Optimization of energy~levels} \label{sec:lev_opt}

The optimized energy values of levels involved in transitions were determined from observed wavelength data. In this regard, a least-squares level optimization was performed using the LOPT code \citep{Kramida2011-lopt}. To optimize the energy levels of the Kr VI spectrum, we have adopted similar methods and processes outlined in the earlier works \citep{Haris2014-Sn2, Kramida2013-crit-Ev, Kramida2013-In2, Kramida2013-Ag2}. Although several iterations of level optimization were made for the spectrum, its initial results were obtained using the Kr VI spectral lines \citep{Tauheed_1990_Kr-VI, Pagan_1995_Kr6_exp, Pagan_1996_Kr6_os}, which were included in the compilation of \citet{Saloman_2007_Kr_compil}. The measurement uncertainties used were their originally reported values (see Section \ref{sec:measur_P96}), and we noticed no inconsistency between observed and Ritz wavelengths. Accordingly, we obtained energy levels and Ritz wavelengths with their estimated uncertainties for the Kr VI spectrum described in Saloman's compilation. In general, it was found that all excited energy levels are accurate within 1 to 2 cm$^{-1}$. On the other hand, in Saloman's compilation, all excited levels of Kr VI were reported with a constant uncertainty of 5 cm$^{-1}$, due to which its Ritz wavelengths are imprecise in the current NIST ASD \citep{nist}. Since our observation supports its several previously known lines with improved uncertainties (see Section \ref{sec:measur_P96}) and additional systematics were also found in the Pagan et al. \citep{Pagan_1995_Kr6_exp, Pagan_1996_Kr6_os} measured wavelengths (see details below), a rigorous optimization of level is required for Kr VI spectrum by combining all accurate and precise measurements. \\

As explained in the end of Section \ref{sec:measur_P96}, though no systematic differences between two measurements (Pagan et al.'s \citep{Pagan_1995_Kr6_exp, Pagan_1996_Kr6_os} vs this work) were noticed when plotted against wavelengths; but their standard deviation (SD) agrees within a doubled figure of the original uncertainties claimed by Pagan et al. \citep{Pagan_1995_Kr6_exp, Pagan_1996_Kr6_os}. However, when the above differences were plotted against their corresponding upper energy levels, a clear systematic increasing trend of their differences or shifts was noticed. We ascribe this to the Stark shifts, which could possibly be present in the theta-pinch type of plasma light sources \citep{ Rocco_1986_stark_shift_Xe, Rocco_1988_Stark_shift_Kr_Xe, Rocco_1989_Stark_shift_Kr_Xe}. Stark broadening and shift data are a prerequisite for several astrophysical and laboratory plasma problems and applications \citep{Brechot_2021_Stark_Br5_Kr5-7}; however, determining Stark broadening parameters is beyond the scope of this work. The only factor that concerns us about determining energy levels requires spectral wavelength data from Stark-free environments or sources \citep{Haris_2022_C2}. Thus, Pagan et al.'s original measurement uncertainties, as indicated by our comparison, were set to 20 m\AA~for unperturbed lines. In the final optimization, the reported wavelengths of Pagan et al. were replaced with our values for all observed features in the spectrogram, and for unobserved features, those were retained at Pagan et al.'s values with their doubled uncertainties. The forbidden transition within the ground 4s$^2$4p $^2$P$^{\circ}$ term with J = 1/2 -- 3/2 was observed in the spectra of many high-ionization nebulae at 12330~\AA~by \citet{Sterling_2017_Kr6_PNe}. Our investigation shows that the observed feature had a linewidth of about 8~\AA, widened due to the blending of other lines reported by Sterling et al. Therefore, the uncertainty of the line was taken as half of its linewidth, and the same was inserted into the level optimization scheme. By this, a total of 169 unique lines classifying 175 transitions were included in the final LOPT to optimize 52 levels, and their results were tabulated in Tables \ref{tab:lin_KrVI} \& \ref{tab:lev_KrVI}. \\ 

It is worth mentioning that \citet{Farias_2011_Kr-VI} observations were excluded from the above-mentioned final LOPT scheme due to several flaws in their measurement and/or limitations of our present observation to sufficiently support the same (see details in Section \ref{sec:measur_F11}). \\
 
\subsection{Theoretical calculations}\label{sec:theo_HFR} 

To support the experimental findings of the present work, theoretical calculations were performed using a pseudo-relativistic Hartree–Fock (HFR) method with a superposition of interacting configurations. These calculations were carried out using Cowan’s suite of codes \citep{Cowan_1981_book}. We used the Windows-based version of the Cowan codes~\citep{Kramida_cowan-code}. The computational approach followed in this work is based on the procedure described in reference \citet{Cowan_1981_book}. In this study, we considered two sets of atomic models, given in Table \ref{tab:List_of_config}, with varying numbers of interacting configurations. The model HFR-A has the same number of interacting configurations as used by~\citet{Rauch_2016_Kr-VI} for Kr VI work, whereas the model HFR-B includes a substantially larger number of interacting configurations. Our HFR-B model is superior to the HFR-A model, not only in terms of the number of configurations, but also because it accounts for the prevailing large valence-valence (VV) electron correlations for the sets of configurations studied in this work (see ref. \citep{Marcinek_1993_os_ga-like} for additional details regarding the electron correlation). In the theoretical model of \citet{Rauch_2016_Kr-VI}, a core-polarization (CPOL) term was added to their HFR model to account for the core-valence (CV) electronic interactions. Thus, a systematic approach will be helpful to us to estimate the effect of these models on the computed transition probabilities or gA-values of Kr VI. \\

%%%%%%%% begin of %%%%%%%%%%%%
%%%%%%%% configurations Table %%%%%%%%%%%%

\begin{table*}[ht] 
\centering
\setlength{\tabcolsep}{12pt}
\begin{scriptsize}
\begin{threeparttable}
\caption{Configurations used in HFR models of Kr VI.}~\label{tab:List_of_config}
\begin{tabular}{lccl}
\hline
{Odd Parity} & & & {Even Parity} \\
\hline
\multicolumn{4}{c}{Model:~HFR-A} \\
\hline
4s$^{2}n\ell$ (n $\le$ 6, $\ell = $ p, f) & & & 
4s$^{2}n\ell$ (n $\le$ 6, $\ell = $ s, d) \\
4p$^3$, 4s4p$n\ell$ (n $\le$ 6, $\ell = $s, d) & & & 4s4p$^2$, 4s4p$n\ell$ (n $\le$ 6, $\ell = $ p, f) \\
4p$^2$4f, 4d$^2$4f & & & 4p$^2$(4d+5s) \\
4p(4d$^2$+4f$^2$) & & & 4s(4d$^2$+4f$^2$) \\
4s4d4f & & & 4s4d5s \\
....... & & & ....... \\
No. of Levels~\textsuperscript{a} = 40\{17\} & & & No. of Levels~\textsuperscript{a} = 12\{9\} \\
SD = 256 cm$^{-1}$ & & & SD = 71 cm$^{-1}$ \\
\hline
\multicolumn{4}{c}{Model:~HFR-B} \\
\hline
4s$^{2}n\ell$ (n $\le$ 10, $\ell = $ p, f; n $\le$ 8, $\ell = $ h) & & & 4s$^{2}n\ell$ (n $\le$ 10, $\ell = $ s, d) \\
4p$^3$, 4s4p$n\ell$ (n $\le$ 7, $\ell = $s, d) & & & 4s4p$^2$, 4s4p$n\ell$ (n $\le$ 8, $\ell = $ p; n $\le$ 6, $\ell = $ f) \\
4p$^{2}n\ell$ (n $\le$ 10, $\ell = $ p, f; n $\le$ 8, $\ell = $ h) & & & 4p$^{2}n\ell$ (n $\le$ 10, $\ell = $ s, d) \\
4p(4d$^2$+4f$^2$+5s$^2$+5p$^2$), 4d$^2$(4f+5p) & & & 4s(4d$^2$+4f$^2$) \\
4s4d(4f+5p+5f), 4s4f(5s+5d), 4s5s(5p+5f) & & & 4s4d5s \\
4p4d(5s+5d), 4d4f(5s+5d), 4f$^2$5p & & & \\
....... & & & ....... \\
No. of Levels~\textsuperscript{a} = 40\{17\} & & & No. of Levels~\textsuperscript{a} = 12\{9\} \\
SD = 248 cm$^{-1}$ & & & SD = 70 cm$^{-1}$ \\
\hline
\end{tabular}
\begin{tablenotes}
\item[a] Total number of known levels and free parameters in the least-squares fitting (LSF), the latter quantity is given in parentheses.
\end{tablenotes}
\end{threeparttable}
\end{scriptsize}
\end{table*}

In both models, the initial values of Slater's parameters were kept at 85\% of the HFR-value for the $F^{k}$, 75\% for the $G^{k}$, 75\% for the $R^{k}$, and~the $E_{av}$ and $\zeta_{n, l}$ parameters were fixed at 100\% of their HFR-values. A parametric LSF adjustment module inherent to the Cowan suite of codes was used to minimize the differences between the calculated and observed energy levels. The quality of the LSF was judged using its SDs. In all LSF calculations, we mainly considered the known energy levels listed in Table \ref{tab:lev_KrVI}, which were either previously reported by \citep{Pagan_1995_Kr6_exp, Pagan_1996_Kr6_os, Saloman_2007_Kr_compil} or newly established by us. Since we could not confirm \citet{Farias_2011_Kr-VI} observations, their reported energy levels were not explicitly included in these LSF calculations. However, their effects on gA-values for transitions shown in Table \ref{tab:lin_KrVI} were made using separate LSF calculations. \\

In the LSF calculations, similar Slater parameters were generally grouped and linked to obtain their optimized values, except for a few parameters, which were kept fixed for better fitting stability. The standard deviation (SD) of fit is given in Table \ref{tab:List_of_config} along with the total number of known energy levels and free parameters used in the fitting, shown in curly brackets. The fitted LSF parameters were used to generate the radiative transition parameters, including gA-values. The gA-values obtained from the HFR-B model, along with their cancellation factors (CFs), are given in column 7 of Table \ref{tab:lin_KrVI}, and the LS percentage compositions are presented in columns 4 to 8 of Table \ref{tab:lev_KrVI}. The LS assignments of the levels reported by Pagan et al. \citep{Pagan_1995_Kr6_exp} were found to be good without much ambiguity in our extensive calculation (HFR-B model). Nonetheless, the presence of strong configuration interaction, prevailing between the levels of odd parity configurations such as $4p^3$, 4s4p4d, 4s4p5s, and 4s$^2$4f, dropped down the LS purity of several levels to below 50\% (see Table \ref{tab:lev_KrVI}). In the subsection below, we discuss one of the main objectives of the present work regarding the evaluation of computed gA-values. \\

\subsubsection{Transition probabilities and their uncertainty estimates}\label{sec:tp_eval}

\begin{figure*}[ht!]
\centering
\advance\leftskip-2cm
\advance\rightskip-2cm
\includegraphics[scale=0.50]{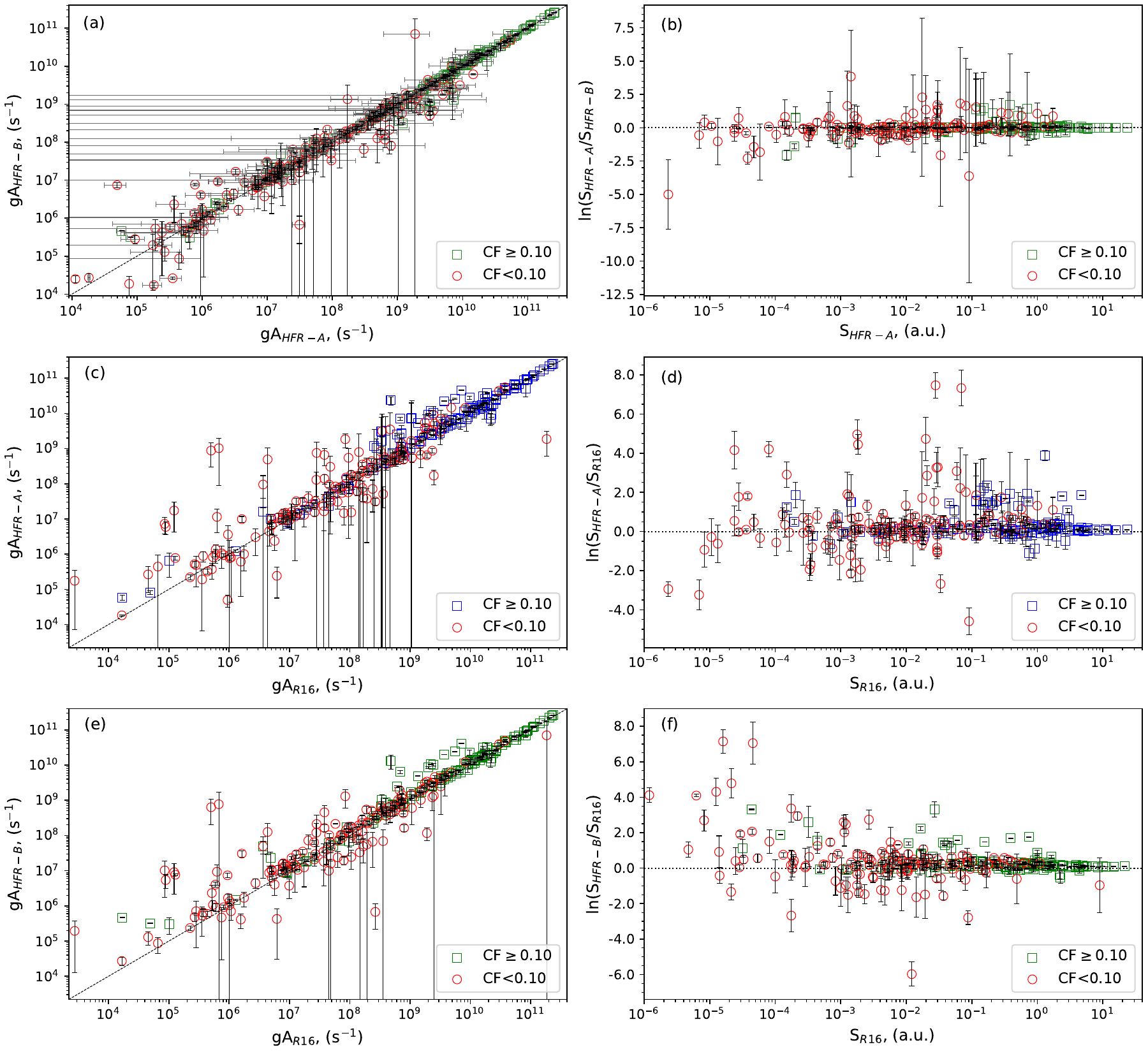}
\caption{Comparison of gA-values (a,c,e) and S-values (b,d,f) obtained from our HFR-A and HFR-B models with those of the HFR+CPOL model by R16---\citet{Rauch_2016_Kr-VI}. The data sets were divided into two groups: CF unaffected (in squares) and affected (in circles) transitions. The error bars represent the internal uncertainties of the quantity involved, which we computed by performing several sets of calculations with varying parameters (see text).} \label{fig:TP_comp_ABR}
\end{figure*}

In general, two types of schemes were employed for comparison and evaluation of gA-values or line strengths (S-values) from two different calculations: (i) a qualitative scatter plot analysis for either gA- or S-values and (ii) a quantitative method based on the uncertainty estimation of $d$S = $ln$(S$_1$/S$_2$) as a function of line-strengths for a set of transitions with similar accuracy \citep{Haris2014-Sn2, Kramida2013-crit-Ev, Kramida2013-In2, Kramida2013-Ag2, Kramida_2024_Evalu}. Before making the comparisons, several sets of calculations were performed by varying the parameters of both HFR-A (9 sets) and HFR-B (6 sets) models. This enabled us to compute the average gA-values or S-values for the transitions listed in Table \ref{tab:lin_KrVI} with their SDs, the latter could be taken as an internal uncertainty estimator for each gA-value or S-value. The comparison schemes for all three data sets -- from the present HFR-A and HFR-B models and the HFR+CPOL model used by \citet{Rauch_2016_Kr-VI} -- were illustrated in Figure~\ref{fig:TP_comp_ABR}. In general, the correlation between the two compared quantities was satisfactory. The reliability of gA-values or S-values is inherently shown by their absolute cancellation factor (|CF|) for the models. In the HFR framework, generally, gA-values or S-values of the transitions with |CF|~$\ge$~0.10 are considered reliable (see details in ref.~\citep{Cowan_1981_book}). Our comparisons reflect that the cancellation affected transitions with |CF|<0.10 have considerable disagreement, and in fact, their SDs (i.e., those with larger error bars in all panels of Figure \ref{fig:TP_comp_ABR}) were also larger. Most of these transitions have smaller S-values. Based on $d$S for a set of transitions with similar accuracy, the uncertainty evaluation is carried out for all three cases and is summarized below. \\

The evaluations started with comparisons of S-values from HFR-A and HFR-B models (see top right panel of Figure~\ref{fig:TP_comp_ABR}) and their disagreement was settled within 50\% for CF unaffected strong lines with average line strength S$_{av} = 1.54$ a.u. (atomic units) and the CF-affected lines with S$_{av} = 0.09$ a.u. disagreed within 140\%. It is evident from this Figure that the uncertainty not only depends on the CF but also varies with the S-values. However, we did not perform the S-value based evaluations for this comparison scheme, because both HFR-A and HFR-B models account only for configuration interactions, which are VV types. Whereas the HFR+CPOL model implemented by \citet{Rauch_2016_Kr-VI} addressed the CV electronic interactions. Therefore, the remaining comparisons of our HFR-A and HFR-B models were made with respect to HFR+CPOL data sets given by \citet{Rauch_2016_Kr-VI}~(hereafter called HFR+CPOL), and the same were illustrated in the panels (c,d) \& (e,f) of Figure \ref{fig:TP_comp_ABR}. The gross disagreement between the data sets from HFR-A and HFR+CPOL models was 2.1 $dex$ and 2.6 $dex$ for CF unaffected strong and CF affected weak lines, respectively. Their further grouping, based on S-values of similar accuracy, was the following: (i) the CF unaffected lines with S~$\ge 0.10$ a.u. were settled within 25\% and 35\% for the remaining weak lines and (ii) CF affected lines with S~$\in[0.003, 9)$~a.u. disagreed within 50\% and the remaining weak lines have about three order-of-magnitude deviation. Finally, the data sets from our superior HFR-B model were compared with those from the HFR+CPOL model. The evaluated uncertainties were the following:- the strong CF unaffected lines with S~$\ge~1$ a.u. have an uncertainty of 17\%, those with S~$\in[0.04, 1)$~a.u. were uncertain by 24\%, and 36\% for the remaining set of lines. The CF affected lines with S~$\ge0.01$~a.u. have 44\% uncertainty, the lines in S~$\in[0.002, 0.01)$~a.u. were uncertain by 50\%, and the rest of the weak lines have large scatters and differ by three orders of magnitude. It should be noted that the gA-values or S-values of our HFR-B data set have their internal uncertainties, stemming from the different sets of calculations performed with varying parameters. The error bars in the panel (f) of Figure \ref{fig:TP_comp_ABR} represent the same. Therefore, to derive the final uncertainty estimates for each transition listed in Table \ref{tab:lin_KrVI}, the above-mentioned S-value based uncertainties (for HFR-B versus HFR+CPOL) were combined in quadrature with their internal uncertainties. Consequently, all lines given in Table \ref{tab:lin_KrVI} were provided with gA-values, |CF|-values (both from the HFR-B calculations), and uncertainty codes. The uncertainty codes are C+ types with an accuracy $\le18\%$, C with $\le25\%$, D+ with $\le40\%$, D with $\le50\%$, and the E types with an accuracy $>50\%$. All these uncertainty estimates must be taken as statistical due to the scarcity of accurate experimental radiative lifetime data for the Kr VI spectrum. \\

It is worth mentioning that \citet{Pagan_1996_Kr6_os} also provided the oscillator strength using the HFR method implemented in the Cowan codes to support their observations. We computed their corresponding S-values, and comparisons were made with their competing data sets from the present HFR-A and HFR-B models and those of HFR+CPOL data sets~\citep{Rauch_2016_Kr-VI}. Their gross disagreements are reasonable and satisfactory within two orders of magnitude. However, their further $d$S-based evaluations were not worth mentioning here because the HFR model used by \citet{Pagan_1996_Kr6_os} is inferior to all other models described in this work. In the recent work, \citet{Rauch_2016_Kr-VI} showed the comparison of log gf-values of selected Kr VI lines relevant to the white dwarf (hot DO-type) RE 0503--289. A total of fourteen Kr VI lines were used in the abundance determination of krypton in RE 0503--289 (see Table 3 of ref. \citep{Rauch_2016_Kr-VI}). We summarized their detailed comparison in Table \ref{tab:lines_in_WD} with |CF|-values and uncertainty estimates. \\

\begin{table*}[ht] 
\begin{threeparttable}
\centering
\begin{scriptsize}
\caption{Comparison of log gF values of Kr VI lines detected in the DO-type white dwarf RE 0503--289.} ~\label{tab:lines_in_WD}
\begin{tabular}{lcclcccc}
\hline \hline
 & $\lambda_{Obs.}$ & \multicolumn{4}{c}{log gf from literature and HFR-B of this work} & Unc.$^a$ & \\
\cline{3-6}
Transition & (\AA) & P96~\citep{Pagan_1996_Kr6_os} & R16~\citep{Rauch_2016_Kr-VI} & TW & |CF| & (\%) & Acc.$^b$ \\ 
\hline
4s4p$^{2}$~$^{2}$S$_{1/2}$-4p$^{3}$~$^{4}$S$^{\circ}_{3/2}$ & 919.936 & -1.966 & -2.22 & -2.085 & 0.05 & 37(44) & E \\
4s$^{2}$4p~$^{2}$P$^{\circ}_{1/2}$-4s4p$^{2}$~$^{4}$P$_{1/2}$ & 927.334 & -2.420 & -2.50 & -2.501 & 0.51 & 21(36) & D+ \\
4s$^{2}$4p~$^{2}$P$^{\circ}_{3/2}$-4s4p$^{2}$~$^{4}$P$_{5/2}$ & 931.335 & -2.005 & -2.05 & -2.031 & 0.20 & 20(36) & D+ \\
4s4p$^{2}$~$^{2}$S$_{1/2}$-4p$^{3}$~$^{2}$D$^{\circ}_{3/2}$ & 944.039 & -0.959 & -1.11 & -1.018 & 0.11 & 19(44) & D \\
4s$^{2}$4d~$^{2}$D$_{3/2}$-4s$^{2}$5p~$^{2}$P$^{\circ}_{3/2}$ & 956.622 & -0.974 & -0.97 & -0.945 & 0.64 & 1(24) & C \\
4s$^{2}$4d~$^{2}$D$_{5/2}$-4s$^{2}$5p~$^{2}$P$^{\circ}_{3/2}$ & 965.103 & -0.019 & -0.02 & 0.007 & 0.68 & 1(17) & C+ \\
4s$^{2}$4p~$^{2}$P$^{\circ}_{3/2}$-4s4p$^{2}$~$^{4}$P$_{3/2}$ & 970.083 & -2.731 & -2.76 & -2.751 & 0.37 & 19(36) & D+ \\
4s$^{2}$4d~$^{2}$D$_{3/2}$-4s$^{2}$5p~$^{2}$P$^{\circ}_{1/2}$ & 980.412 & -0.278 & -0.28 & -0.252 & 0.69 & 1(17) & C+ \\
4s$^{2}$4p~$^{2}$P$^{\circ}_{3/2}$-4s4p$^{2}$~$^{4}$P$_{1/2}$ & 1002.732 & -2.746 & -2.88 & -2.865 & 0.16 & 15(36) & D+ \\
4s$^{2}$5s~$^{2}$S$_{1/2}$-4s4p($^{3}$P$^{\circ}$)4d~$^{2}$P$^{\circ}_{3/2}$ & 1011.152 & -1.826 & -1.84 & -1.736 & 0.38 & 3(24) & C \\
4s4p$^{2}$~$^{2}$P$_{1/2}$-4p$^{3}$~$^{4}$S$^{\circ}_{3/2}$ & 1015.761 & -1.580 & -1.83 & -1.563 & 0.06 & 13(44) & D \\
4s4p$^{2}$~$^{2}$P$_{1/2}$-4p$^{3}$~$^{2}$D$^{\circ}_{3/2}$ & 1045.227 & -0.823 & -0.98 & -0.773 & 0.08 & 12(44) & D \\
4s4p$^{2}$~$^{2}$P$_{3/2}$-4p$^{3}$~$^{4}$S$^{\circ}_{3/2}$ & 1052.963 & -1.561 & -1.72 & -1.584 & 0.12 & 7(44) & D+ \\
4s4p$^{2}$~$^{2}$P$_{3/2}$-4p$^{3}$~$^{2}$D$^{\circ}_{5/2}$ & 1061.046 & -0.409 & -0.58 & -0.393 & 0.12 & 8(24) & C \\
\hline
\end{tabular}
\begin{tablenotes}
\item[a] Uncertainty components for given log gf-value: first figure corresponds to the SD, which was derived by performing several sets of calculations with varying parameters, and the quantity given in parentheses was obtained from the $d$S comparison employed for the HFR-B and HFR+CPOL data sets (see text). 
\item[b] The accuracy code in the last column corresponds to the combined figure of the uncertainty components from the preceding column (see text) (see text).
\end{tablenotes}
\end{scriptsize}
\end{threeparttable}
\end{table*}

The radiative transition parameters of the astrophysically important forbidden line within the ground term 4s$^2$4p $^2$P$^{\circ}$ with J = 1/2 -- 3/2 were available from the following literature sources \citep{Biemont_1987_tp_fb, Ali_1997_fine-str_Ga-like, Safronova_2006_4p_Ga-like, Charro_2008_tp_Ga-like}, which were essentially computed by using different theoretical methods. We presented all those in Table \ref{tab:FB_line_TP} together with the results obtained from our HFR-B model. The following computational methods were used: (i) \citet{Ali_1997_fine-str_Ga-like} employed multi-configuration Dirac-Fock (MCDF), (ii) all-order relativistic many-body perturbation theory (RMBPT) by \citet{Safronova_2006_4p_Ga-like}, (iii) the relativistic quantum defect orbital (RQDO) method was used by \citet{Charro_2008_tp_Ga-like}, and (iv) \citep{Biemont_1987_tp_fb} used the same HFR method with parametric LSF adjustments implemented in Cowan codes (see Section \ref{sec:theo_HFR}). Though our calculations contain large sets of interacting configurations from the HFR-B model, our obtained transition rates (A-values) are almost the same as those from \citet{Biemont_1987_tp_fb} (see Table \ref{tab:FB_line_TP}). The most accurate results were expected from the RMBPT calculations with all-order single-double (SD) method, which is often referred to as the Z$^{SD}$ \citep{Safronova_2006_4p_Ga-like}. These authors evaluated the reduced matrix elements of the magnetic-dipole (M1) and electric-quadrupole (E2) transitions in many orders of the RMBPT calculations, including the all-order Z$^{SD}$ and the radiative correction term Z$^{rad}$ for the M1 transition. We computed their corresponding A-values, and the obtained figures are presented in Table \ref{tab:FB_line_TP}. It is evident from this table that all computed values, except those of the RQDO method by \citet{Charro_2008_tp_Ga-like} for both M1 and E2 transitions, agree within an SD of $\approx1\%$. The branching fraction of M1 and E2 transitions was also given in the table, and the M1 transition is the dominant one that defines the radiative lifetime of the upper level. We recommend the Z$^{SD}$ results with the above-mentioned uncertainty estimates. Compared to others, \citet{Charro_2008_tp_Ga-like} published results disagree by more than 50\% and 10\% for M1 and E2, respectively. Therefore, \citet{Charro_2008_tp_Ga-like} results may be disregarded.

\begin{table}[h]
\centering
\begin{threeparttable}
\caption{Radiative transition parameters for 4s$^2$4p $^2$P$^{\circ}$ J = 1/2 -- 3/2 line} \label{tab:FB_line_TP}
\begin{tabular}{lcclccc}
\hline \hline
\multirow{2}{*}{Theoretical method$^a$} & \multicolumn{2}{c}{M1 transition} &  & \multicolumn{2}{c}{E2 transition} & \multirow{2}{*}{$\tau$, (s)}$^c$ \\
\cline{2-3} \cline{5-6}
 & A, (s$^{-1}$) & BF$^b$ & & A, (s$^{-1}$) & BF$^b$ & \\
\hline
HFR+LSF; B87 & 4.82 & 0.99885 &  & 0.00555 & 0.00115 & 0.2072 \\
MCDF; A97 & 4.79 & 0.99890 &  & 0.00528 & 0.00110 & 0.2085 \\
ZDT RMBPT; S06 & 4.746 & 0.99893 &  & 0.00509 & 0.00107 & 0.2105 \\
RQDO; C08 & 2.72 & 0.99837 &  & 0.00444 & 0.00163 & 0.3670 \\
HFR+LSF; TW & 4.81 & 0.99884 &  & 0.00561 & 0.00116 & 0.2075 \\
\hline
\end{tabular}
\begin{tablenotes}
\item[a] Theoretical method (see text for their explanation) and reference code for A-values of M1 and E2 transitions given in columns 2 \& 4. The reference codes: A97---\citet{Ali_1997_fine-str_Ga-like}, B87---\citet{Biemont_1987_tp_fb}, C08---\citet{Charro_2008_tp_Ga-like}, S06---\citet{Safronova_2006_4p_Ga-like}, TW---this work.
\item[b] Computed branching fractions of M1 and E2 transitions.
\item[c] Theoretical lifetime of the upper level. 
\end{tablenotes}
\end{threeparttable}
\end{table}

\subsection {Intensities of observed lines}
\label{sec:int_model}

For Kr VI lines presented in this work, a Boltzmann plot-based intensity modeling, which was described in detail and successfully applied to several spectra in previous works \citep{Haris2014-Sn2, Kramida2013-crit-Ev, Kramida2013-In2, Kramida2013-Ag2}, was employed to reduce the observed intensities to a common uniform scale. The final modeled line intensities were obtained after removing the wavelength-dependent response function of the instrument, including those of the photographic plates, and/or by rectifying the non-linear response of the photographic plates. The separate modeling was performed for the lines observed on our plates (see Table \ref{tab:lin_KrVI}) and for those originally reported by Pagan et al. \citep{Pagan_1995_Kr6_exp, Pagan_1996_Kr6_os}, and their final effective source excitation temperatures were found to be nearly the same: 7.6 eV and 7.9 eV, respectively. Nonetheless, the fitted parameters from the Boltzmann plot for our observations were selected to scale the modeled intensities of Pagan et al~\citep{Pagan_1995_Kr6_exp, Pagan_1996_Kr6_os}. For common lines observed in both works, their intensities were averaged if they agreed within one order of magnitude. The final modeled intensities of Kr VI reported in this work were given in column 1 of Table \ref{tab:lin_KrVI} with their line characters and remarks (see Table footnotes for additional details). \\ 

\section{Conclusions} \label{sec:conclusion}

In this work, the spectrum of Kr VI has been investigated using krypton's spectrogram recorded on a 3 m NIVS in the wavelength range of 230--2075~{\AA}. We have critically examined and evaluated all previously reported spectral works of this spectrum. For example, the previous observations by \citet{Pagan_1995_Kr6_exp, Pagan_1996_Kr6_os} have been verified by us, whereas those of \citet{Farias_2011_Kr-VI} have not been confirmed. Many sets of extended HFR calculations have been performed with LSF adjustments to support and validate the present experimental findings. As a result, the gA-values or S-values, including the literature data, and their comparisons and evaluations have been performed. Finally, all observed and Ritz wavelengths of Kr VI have been presented with their uncertainties, modeled line intensities and line characters, and with evaluated gA-values.\\ 

% \noindent
% \textbf{Funding Statement}

% This research did not receive any specific grant from funding agencies in the public, commercial, or not-for-profit sectors. \\

% \noindent
% \textbf{Declaration of Competing Interest}

% The authors declare that they have no known competing financial interests or personal relationships that could have appeared to influence the work reported in this paper. \\

% \noindent
% \textbf{CRediT authorship contribution statement} \\

% \textbf{Aftab Alam:} Formal analysis, Validation, Writing - Original Draft. \textbf{Abid Husain:} Conceptualization, Investigation, Formal analysis. Writing - original draft. \textbf{K. Haris:} Conceptualization, Methodology, Formal analysis, Investigation, Validation, Writing - Original Draft, review \& editing, Visualization. \textbf{A. Tauheed:} Investigation, Resources, Supervision. \textbf{S. Jabeen}: Resources. \\
% \\
% \noindent
\textbf{Acknowledgments}

A. Tauheed would like to acknowledge Prof. (late) Y. N. Joshi for his support and hospitality in recording the krypton spectrograms during his visit to the Antigonish lab. We also thank Dr. Alexander Kramida of NIST, Gaithersburg, for his valuable suggestions and discussions. \\

% %Data availability
% \noindent
\textbf{Data Availability}

 The authors confirm that the primary data supporting the findings of this study are available within the article or its supplementary materials. Other auxiliary data supporting this study's findings are available from the corresponding author, [K. H.], upon reasonable request. \\

% \noindent
% \textbf{Supplementary materials}

% {Supplementary material associated with this article can be
% found, in the online version at doi:@-@-@-@-@-@.}

%% The Appendices part is started with the command \appendix; %% appendix sections are then done as normal sections
%\appendix
%\input{appendix}

%Insert \textcolor{red}{Table } 

%% For citations use: 
%% \citet{<label>} ==> Jones et al. [21]
%% \citep{<label>} ==> [21]
%%

%\begin{thebibliography}{00}
\bibliographystyle{elsarticle-num-names} 
\bibliography{Aab-Ref.bib}

\end{document}